\documentclass[10pt,a4paper,twocolumn,twoside,amsmath,showkeys,aps,pra]{revtex4-1}
\usepackage{natbib}
\usepackage[english]{babel}
\usepackage[utf8]{inputenc}
\usepackage[colorinlistoftodos, color=green!40, prependcaption]{todonotes}
\usepackage{tabularx}
\usepackage{ulem}
\usepackage{graphics}
\usepackage{xcolor,soul}
\usepackage{wrapfig}
\usepackage[rightcaption]{sidecap}
\def\ii{{\rm i}}
\def\dd{{\rm d}}
\def\ee{{\rm e}}

\begin{document}
\title{Wigner Function for Harmonic Oscillator\\ and The Classical Limit}

\author{Jan Mostowski, Joanna Pietraszewicz}
    \email{pietras@ifpan.edu.pl}
    \affiliation{Institute of Physics, Al. Lotnikow 32/46, PL-02668 Warsaw, Poland}


\begin{abstract}
The Wigner function is a quantum analogue of the classical joined distribution of position and momentum. As such is should be a good tool to study quantum--classical correspondence. In this paper, the classical limit of the Wigner function is shown using the quantum harmonic oscillator as an~example. The Wigner function is found exactly for all states. The semi-classical wavefunctions for highly excited states are used as the approach to the classical limit. Therefore, one can found the classical limit of the Wigner function for highly excited states and shown that it gives the classical microcanonical ensemble. 
\end{abstract}

\keywords{Harmonic oscillator, Wigner function, Semiclassical approximation\vspace*{0.9cm}}

\maketitle
\section{Introduction}

Relation between classical and quantum mechanics is one of important elements of any standard course in basic quantum mechanics. Unfortunately, it is not an easy subject to discuss. 

A classical mechanical system is usually described by Newton's equations. Sometimes more refined equations of motion are used, e.g. Hamilton's equations or the Hamilton--Jacobi equation, etc. All these approaches rely on trajectories in the configuration space or the phase space, as the main elements. Relation between such trajectories and wavefunctions in quantum mechanics is not clear. 

Let us focus for a moment on the alternative approach to classical mechanics, based on the density of trajectories in the phase space. The density satisfies Liouville's equation which replaces equations of motion. Such approach is particularly applicable, when the rules of classical mechanics are used in statistical physics --- density of trajectories in the phase space become a~natural tool. 

The standard approach to quantum mechanics starts with wavefunctions that satisfy the Schr\"odinger equation. Examples of stationary wavefunctions in some systems, like the harmonic oscillator, one electron atom etc. are discussed as typical issue. However nothing else appears that would help explain the relation between these wavefunctions and the classical trajectories.

The semiclassical approximation in which the phase of the wavefunction is related to the classical action, for most of the students turns out not to be sufficient to fully realize the relation. Other vague statements found in some textbooks, that in the limit of small $\hbar$ compared to other characteristic actions, the quantum physics reaches the classical limit, is also far from clear. Students often ask the question what is the classical limit, i.e., when $\hbar \rightarrow 0$, of the wavefunction. Such limit does not exist, therefore the classical limit remains a kind of mystery in standard textbooks. 

It is difficult to find traces of classical trajectories in the quantum stationary wavefunctions. A question arises if quantum mechanics can be formulated in a way that would be similar to the classical approach based on the density in the phase space. Such linking is indeed possible, the necessary tool was introduced by Eugene Wigner and is now called the Wigner function. It is the quantum analogue of the joined distribution of position and momentum. Description of a quantum systems with the help of the Wigner function, is therefore, the analogue to a description of the classical system in the phase space. In the present paper we will focus on a simple example of a one dimensional harmonic oscillator to illustrate this. In addition, we will discuss in detail how the classical limit is reached. 

The Wigner function allows to calculate average values of symmetric functions of position and momentum, i.e., such functions that do not depend on the order of these quantities. For such function the Wigner function can be treated as if it was a joined probability distribution. 

The Wigner function depends on the quantum states, for many states it is not positive and therefore it is not a probability distribution. We will show, however, how does this function behave in the classical limit.  This provides a bridge between the quantum and classical mechanics and leads to better understanding of quantum mechanics.

In this paper we will use example of a harmonic oscillator. This example is simple enough to find exact wavefunction and their time evolution. We will calculate the Wigner function for coherent states and for energy eigenstates to discuss the classical limit. In this way we will explain the role of the semiclassical approximation in discussing this limit.  

A lot of attention has been paid in recent years to the classical limit in the framework of open systems, see e.g.~\cite{zurek}. We will not go into his more advanced topic, but rather concentrate on a simple approach of pure states. 

\section{Quantum harmonic oscillator} 

One dimensional harmonic oscillator in  quantum mechanic, see e.g. \cite{Landau}, is described by the Hamiltonian
\begin{equation}
    H=\frac{p^2}{2m}+\frac{m\omega^2}{2}\, x^2,
\end{equation}
where $p$ denotes the momentum operator and $x$ --- the position operator. The particle undergoes oscillations with the frequency $\omega$. Position and momentum operators satisfy the canonical commutation relations $\left[p,x\right]=-i\hbar$. The energy levels are $E_n=\hbar\omega\,  (n+\frac{1}{2})$, with $n=0,1,\dots$, while the~energy eigenfunctions are 
\begin{equation}
\psi_n(x)=\frac{1}{\sqrt{2^n\,n!\,a\,\sqrt{\pi}}}\, e^{-x^2/2a^2}\, H_n\left(x\right),
\label{energy eigenstates}
\end{equation}
with $H_n(x)$ as Hermite polynomial and the~characteristic length scale $a=\sqrt{\frac{\hbar}{m\omega}}$.

This, in fact, is the standard textbook material. The~only elements worth commenting is that none of these eigenfunctions has a defined $ \hbar \rightarrow 0 $ limit, and hence it is not a proper object to study the classical limit.

As a remedy, coherent states $\psi_z(x)$ are often used \cite{Klauder, Glauber} to describe the quantum--classical correspondence. These quantum states are defined for all complex numbers $z$, and are given by a linear superposition of the energy eigenstates:
\begin{equation}
\psi_z(x)=\sum_n e^{-|z|^2/2}\, \frac{z^n}{\sqrt{n!}}\, \psi_n(x).
\label{coherent}
\end{equation}
The explicit form can be given by performing the summation in~Eq.~(\ref{coherent}). The result is:
\begin{equation}
\psi_z(x)=\frac{1}{\sqrt{a\sqrt{\pi}}}\, \ee^{-\big(|z|^2+z^2\big)/2}\, \ee^{-x^2/2a^2}\ \ee^{\sqrt{2}z\, x/a}.
\label{coherent_explicit}
\end{equation}

The coherent states are well concentrated around \mbox{$x=a(z+z^*)/\sqrt{2}$,} and their Fourier transforms which give the momentum distribution, are well concentrated around $p=\ii \hbar(z^*-z)/\sqrt{2}a$. The time evolution of the coherent states is very simple: each state labeled by $z$ remains coherent with a different value of the parameter $z$ due to its evolution equation $z(t)=z\exp(-\ii\, \omega t)$. We~refer to other papers on the subject, see e.g. \cite{Schleich}. 

The above features of the coherent states make them good candidates to discuss the quantum--classical correspondence. The discussion on this point is given in the next section.

\section{Wigner function }
The Wigner function $W(X,P)$ is the quantum analogue of the joint probability distribution of position and momentum. For a given state $\psi(x)$ it is defined as follows:
\begin{eqnarray}
    &&W(X,P)=\\ 
    && \int\,  \frac{\dd{}s}{2\pi\hbar}\ \psi{^*}\left(X-s/2\right)\, \exp\left(-\ii\, P s/\hbar\right)\, \psi\left(X+s/2\right)\nonumber\label{Wigner}.
\end{eqnarray}
The Wigner function depends on $X$ and $P$ --- position and momentum. It can be easily checked that the integral over $P$ gives the position distribution expressed in the standard way by the modulus squared of the wavefunction, i.e.,
\begin{equation}
    \int \dd{}P\, W(X,P)=|\psi(X)|^2.
\end{equation}
In a similar way, the integral over $X$ gives the momentum distribution, i.e.,
\begin{equation}
    \int \dd{}X\, W(X,P) = |\widetilde{\psi}(P)|^2
\end{equation}
where $\widetilde{\psi}$ is the Fourier transform of $\psi$.

The Wigner function was introduced by~E.~Wigner~\cite{Wigner2}. A very nice introduction to this function can be found in~\cite{Case}, for more on the subject see e.g.~\mbox{\cite{textbook, Schleich, Hillery84, Lee}.} 

The Wigner function is a natural tool providing a bridge between quantum and classical mechanics. The Wigner can be considered as a kind of joined distribution function of position and momentum. This is because some quantum averages can be calculated in the same way as in classical physics. This applies to quantities that are symmetric functions of position and momentum. Thus, the average value of $xp+px$, i.e., a symmetric function of $x$ and $p$, is given by
\begin{eqnarray}
    \langle xp+px \rangle&=& \int\dd{}x\ \psi^*\left(x\right) \left(xp+px\right) \psi\left(x\right)=\nonumber\\
    &=&2\int \dd{}X\int \dd{}P\  W(X,P)\ XP\, .
\end{eqnarray}
Functions of $x$ and $p$ that are not symmetric can be always reduced to a combination of symmetric functions, e.g. \mbox{$xp= \tfrac{1}{2}\left(xp+px+\ii\hbar\right)$.} It is clear that expectation values of a large class of physical quantities that are functions of position and momentum can be found with the help of the Wigner function used as joined probability distribution of $x$ and $p$.  

For more comprehensive overview on $W(X,P)$ we refer to~\cite{Tatarskij83}, where several cases of (a)~symmetric operators in $x$ and $p$ like $\hat{o}_S=\hat{p}^2\hat{x}+2\, \hat{p}\hat{x}\hat{p}+\hat{x}\hat{p}^2$ and (b)~non-symmetric like $\hat{o}_N=px^2p=\frac{1}{2}\left(\hat{p}^2\hat{x}^2+x^2\hat{p}^2\right)+\hbar^2$, are analyzed. So, in the case of $\hat{o}_N$, there is no classical--quantum correspondence and the average values calculated such way will differ substantially. In turn, in the case of $\hat{o}_S$, there exists a linking path between values $\langle\hat{p}\rangle=p$ and  $\langle\hat{x}\rangle=x$.

Note that in the classical limit all functions of $X$ and $P$ are symmetric --- the quantities $X$ and $P$ commute in this limit. By indicating the quantum operator of such feature of $X$ and $P$, then a joined probability distribution, i.e. $W(X,P)$, becomes a classical distribution of position and momentum. 

\subsection{Examples}

Let us check and discuss examples of the Wigner functions and their classical limit. We start with the ground state given in~Eq.(\ref{energy eigenstates}) for $n=0$. The integration over~$s$ in~Eq.(\ref{Wigner}) amounts to the integration of gaussian functions. The Wigner function turns out to be
\begin{equation}
    W_0(X,P)=\frac{1}{\pi\, \hbar}\ \ee^{-X^2/a^2} \ee^{-P^2 a^2/\hbar^2}. 
\end{equation}
Taking naively the classical limit, i.e., $\hbar\to 0$ we get the following result     
\begin{equation}
     W_0(X,P)\longrightarrow \delta(X)\, \delta(P) 
     \label{ground state limit}
\end{equation}
which is due to the standard mathematical property $\lim_{b\rightarrow 0} \frac{1}{\sqrt{\pi b}}\, \exp(-x^2/b) = \delta(x)$. From the formal point of view, the limit in Eq.(\ref{ground state limit}) is correct, however, it does not tell much about the quantum--classical correspondence.

The next example is the Wigner function for coherent states. The definition given in Eq.~(\ref{Wigner}) leads to a simple integration of gaussian functions, the result can be easily found: 
\begin{eqnarray}
    W_z(X,P)=\frac{1}{\pi\, \hbar}\, \ee^{-(X-X_0)^2/a^2}\ \ee^{-a^2 (P-P_0)^2/\hbar^2}, \nonumber\\
\end{eqnarray}
where $X_0=\tfrac{a}{\sqrt{2}}\, (z^*+z)$ and $P_0=\tfrac{\ii\, \hbar}{  a\sqrt{2}}\, (z^*-z)$. The classical limit, sometimes but incorrectly called ``the limit of small~$\hbar$'' should be taken in such way that the physical quantities of a classical meaning should go their classical counterparts. The simplest case is such, that these physical quantities remain constant. In the present case, the physical quantities having classical meaning for coherent states are average values of position and momentum. These values should stay constant in the classical limit, i.e., when taking $\hbar \rightarrow 0$ and ensuring \mbox{$|z|\rightarrow \infty$}. So the values of $X_0=\sqrt{\frac{\hbar}{2m\omega}}(z+z^*)$ and $P_0=\ii\,  \sqrt{\frac{m\, \omega \textcolor{white}{1} }{2\hbar}}\, (z^*-z)$ are constants. Hence,  
\begin{equation}
    W_z(X,P)\longrightarrow \delta(X-X_0)\ \delta(P-P_0).
\end{equation}
provides interpretation of the Wigner function for coherent states. The real and imaginary parts of the $z$ parameter determine the~average position and momentum of the particle in such a state. For large $|z|$ the Wigner function becomes close to the classical distribution of position and momentum related to the trajectory with the initial position $X_0$ and momentum $P_0$. Thus, let us emphasize once again, that the classical limit should be understood as $|z|\rightarrow \infty$, $\hbar\rightarrow 0$ but $a\, z = const$.

\begin{figure}[b]
    \centering
    \includegraphics[width=9cm]{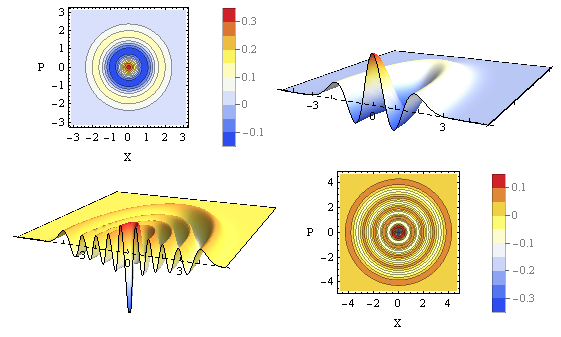}
    \caption{Graphical representation (3D plot and map) of the Wigner function for quantum number $n=2$ (upper panel) and $n=9$ (lower panel) in the phase space of $X,P$. Spatial coordinate is is units of $a$, momentum coordinate --- in units of $\hbar/a$. The number of oscillations which increase with $n$, deserves special attention, as well as the fact that it takes the negative values. }
    \label{fig:fig1}
\end{figure}
\begin{figure}[th]
    \centering   
    \includegraphics[width=7.5cm]{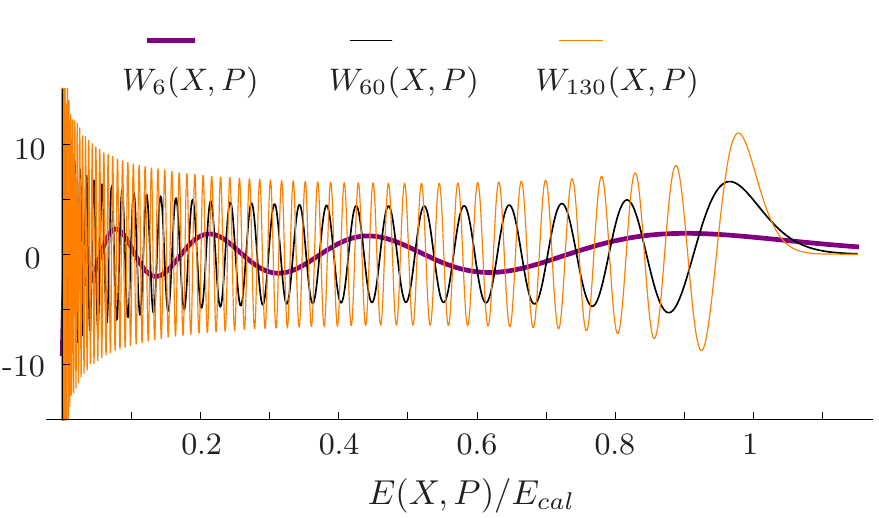}
     \includegraphics[width=7.5cm]{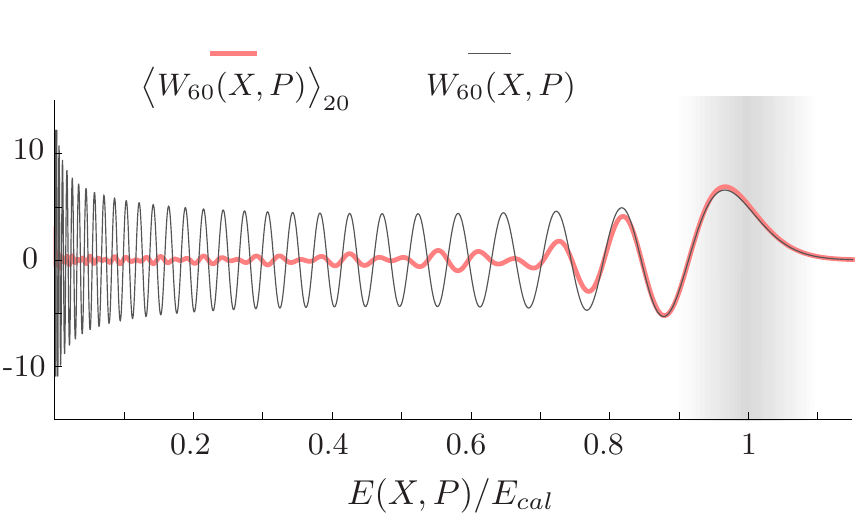}
    \caption{Wigner function as a function of parameter $E(X,P)/E_{clas}$ visualized for large quantum numbers $n$. (top)~thick purple line $n=9$, black line $n=60$, orange line $n=130$. Due to Eq.~(\ref{wignerEn}), it is how the classical limit is realized in the harmonic oscillator system when $ E(X,P)\to E_{clas}$ (shaded area). (bottom) The averaged Wigner function (pink) constructed with $2N_W=20$ individual Wigner functions, i.e., $\langle W_{60}\rangle=\sum_{j=1}^{2N_W} W_{60\pm j}(X,P)$. One can see the mechanism how the neighboring states of similar energy contribute to the classical analogue of Wigner in WKB approximation.}
    \label{fig:fig2}
\end{figure}

We will continue with the discussion of the Wigner function for the energy eigenfunctions (see Eq.~(\ref{energy eigenstates})). We will trace the behaviour of its eigen energy $E$ when goes towards the energy classical variant $E_{clas}$. The Wigner function for excited states with an arbitrary quantum number $n$ is not well known. One can, however, express it in terms of Laguerre polynomials with the argument $\xi=(X/a)^2+(P\,a/\hbar)^2$~\cite{Wang}. Thus, 
\begin{equation}
    W_n\big(\xi(X,P)\big)=\frac{(-1)^n}{\pi\hbar}\, \exp\left(-\xi\right)\, L_n\left(2\xi\right).
    \label{Laguerre}
\end{equation}
The above formula is given e.g. in Wikipedia (with no reference), the derivation is outlined in the Appendix. This form of the Wigner function can be useful, the Laguerre polynomials are build in some applications, like Mathematica, and thus the numerical values can be easily calculated. 

In further discussion we will use the following scaled quantities: $E(X,P)=\frac{m\omega^2}{2} X^2+\frac{1}{2m}P^2$ and $E_{cal}=\hbar\omega n$ and also use the explicit form of $a=\sqrt{\hbar/m\omega}$. One can rewrite $\xi=2n\, \frac{ E(X,P)}{E_{clas}}$ and then 
\begin{eqnarray}
     W_n\big(&&E(X,P)/E_{clas}\big) =\\
    &&\frac{(-1)^n}{\pi}\, \frac{n\, \omega}{E_{clas}}\,  \ee^{-2n \frac{E(X,P)}{E_{clas}}} \, L_n\left( 4n \frac{E(X,P)}{E_{clas}}\right). \nonumber 
    \label{wignerEn}
\end{eqnarray}
This form allows to visualize the Wigner function as a function of one quantity, namely of the energy scaled $\frac{E(X,P)}{E_{clas}}$ for large values of $n$, see~Fig.~\ref{fig:fig2}. If the Wigner function would be close to the classical distribution then these plots should concentrate at $\tfrac{E(X,P)}{E_{clas}}=1$ and be close to zero in the classically forbidden region. As mentioned earlier, this is not the case. The~obtained values oscillate in the whole region of energies from 0 to the classical values of energy $E_{clas}$. The amplitude of these oscillations does not decrease when $n$ increases, on the contrary, it~has a tendency to grow. Moreover, the spatial frequency of these oscillations grows. A peak at the classical value of energy, namely at~$E(X,P)/E_{clas}=1$, emerges for large $n$. As seen in Fig.~(\ref{fig:fig2}), the Wigner function does not have a clear classical limit --- the oscillations in the classically forbidden region remain even in the case of very large $n$.

An interpretation of such behavior of the Wigner function is given in the next section.

\section{Semiclassical approximation}
The standard method to study the quantum-classical correspondence is to use the semiclassical (WKB) approximation. The wavefunction of an energy eigenstate with a large quantum number $n$ can be approximated with the help of the semiclassical (WKB) method~\cite{Landau}, as follows:
\begin{equation}
    \psi_n(x)\simeq \psi^{^{\rm wkb}}_{n}(x)=\frac{A}{\sqrt{p(x)}}\, \sin\left( \frac{1}{\hbar}\mathcal{S}(x)+\mu\right).
    \label{WKB wavefunction}
\end{equation}
The normalization constant is \mbox{$A=\sqrt{4m/T}$,} where $T$ is the period of motion $T=\tfrac{2\pi}{\omega}$, $\mathcal{S}(x)$ is the classical action corresponding to energy \mbox{$E=\hbar \omega (n+\tfrac{1}{2})$} and $p(x)=\frac{{\rm d}S(x)}{{\rm d}x}$. For the~case of harmonic oscillator potential, the Maslov index $\mu=\tfrac{1}{2}$. The classical action $\mathcal{S}(x)$ is given by
\begin{eqnarray}
    \mathcal{S}(x)= \int\limits_0^x \dd{}y\, p(y)=\sqrt{2m}\, \int\limits_0^x \dd{}y\, \sqrt{E-\tfrac{1}{2}\, m \omega^2\, y^2}.\nonumber \\ 
    \label{classical action}
\end{eqnarray}
The lower limit of integration (here taken to be 0) is arbitrary it just changes the value of an arbitrary additive constant. 

This form of the wavefunction is valid in the classically allowed region, but not in the vicinity of the turning points, where $p(x)=0$.

We will now find the Wigner function obtained from the semi-classical wave functions. The expression for the Wigner function takes the form: 
\begin{eqnarray}
  && W_n^{\rm wkb}(X,P)=A^2\int \frac{\dd{}s}{2\pi\hbar }\  {\rm e}^{\ii \,Ps/\hbar} \times \nonumber \\ 
    &&  \qquad\frac{\sin\Big(\frac{1}{\hbar}\mathcal{S}(X-s/2)+\mu \Big)}{\sqrt{p(X-s/2)}}\, 
           \frac{\sin\Big(\frac{1}{\hbar}\mathcal{S}(X+s/2)+\mu\Big)}{\sqrt{p(X+s/2)}},\nonumber \\
    \label{semiwig general}
\end{eqnarray}
The contributions from the vicinity of the turning points and from the classically forbidden region have been neglected. The integration is limited to such values of $s$ that the arguments $X\pm s/2$ are within the classically allowed region, i.e. $x_{min}<X+s/2<x_{max}$ and $x_{min}<X-s/2<x_{max}$, where $x_{min}=-\sqrt{\tfrac{2E}{m\omega^2}}$ and  $x_{max}=\sqrt{\tfrac{2E}{m\omega^2}}$. 

\subsection{Calculations}

\begin{figure}[t!]
    \centering
    \includegraphics[width=8cm]{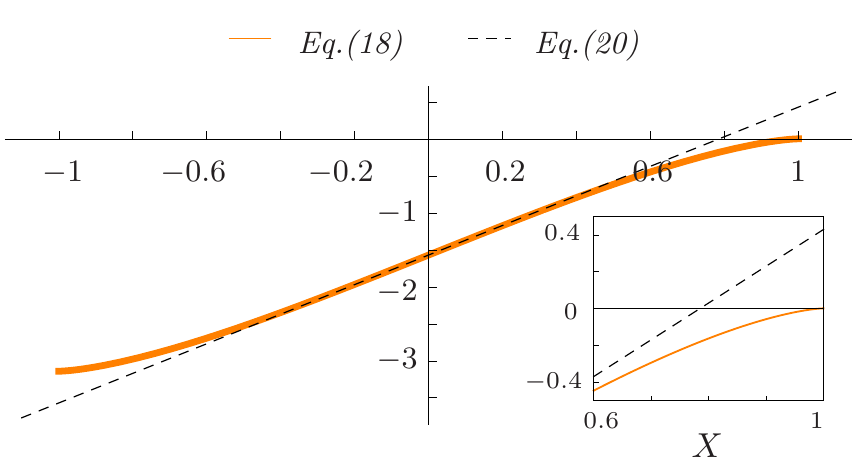}
    \caption{Exact solution given by Eq.(\ref{action_fi}) and the approximated solution Eq.(\ref{approx_action}) indicated by a dashed line. The apparent difference between the lines increasing from the value of $\approx 0.6 x_T$, is clearly seen in the vicinity of the turning point (inset).}
    \label{fig:action}
\end{figure}

In order to proceed we will study the function $\mathcal{S}(X)$ more closely. 
The exact formula is
\begin{equation}
     \mathcal{S}(X)=\frac{E}{\omega} \, \left(\phi+\frac{1}{2}\, \sin(2\phi)+\frac{\pi}{2}\right) 
      \label{action_fi}
\end{equation}
with 
\begin{equation}
\phi=\arcsin\left(\sqrt{\frac{m\omega^2}{2E}}X\right),
\end{equation}
The function $\mathcal{S}(X)$ is plotted in Fig.~\ref{fig:action}. It is seen that the this function can be approximated by a linear function in a broad range of arguments. Therefore, we can expand the action $\mathcal{S}(X\pm s/2)$ up to the linear order in~$s$,
\begin{equation}
    \mathcal{S}(X\pm s/2)=S(X)\pm \frac{p(X)}{2}\, s+\mathcal{O}(s^2)
    \label{approx_action}
\end{equation} 
in the exponent and neglect the $s$ dependence in the denominator of Eq.~(\ref{semiwig general}). This expansion is valid if $s$ is small compared to the scale over which the local de Broglie wavelength changes, i.e.,  $|\frac{1}{p(X)}\frac{\dd{}p(X)}{\dd{}X} s| \ll 1$.

These circumstances allow to express the Wigner as 
\begin{eqnarray}
   && W_n^{\rm wkb}(X,P)=\frac{A^2}{2|p(X)|}\,
   \int\, \frac{\dd{}s}{2\pi\, \hbar}\ \ee^{\ii\, s P/\hbar}\ \times\nonumber \\ 
   && \qquad
   \Big[   \cos\big(s\,p(X)/\hbar\big)-\cos\big(\pi+2\mu+2S(X)/\hbar\big) \Big]. \nonumber\\
    \label{expansion2}
\end{eqnarray}

The integration over $s$ should be taken over the allowed region, for $X>0$ the result is
\begin{eqnarray}
 &&   W^{\rm wkb}(X,P)=\\
     & &\frac{A^2}{4\pi |p(X)|} \bigg(\frac{1}{P+p(X)}\ \sin\left(\tfrac{2 }{\hbar} \left(P+p(X)\right) \left(x_{max}-X\right)\right)\nonumber\\
    &&\qquad +\frac{1}{P-p(X)}\ \sin\left(\tfrac{2}{\hbar}\left(P-p(X)\right)\left(x_{max}-X\right)\right) \bigg)\nonumber \\
  &- &\frac{A^2}{2\pi\, P\, |p(X)|} \sin\left(\tfrac{2}{\hbar}P\left(x_{max}-X\right)\right) \cos\big(\pi+2\mu+\tfrac{2S(X)}{\hbar}\big)\nonumber 
\end{eqnarray}
A similar result is valid for $X<0$. The semi-classical approximation gives a fairly simple expression for the Wigner function. However, this expression does not depend only on $\xi=\frac{X^2}{a^2}+\frac{a^2P^2}{\hbar^2}$ but rather on $X$ and $P$ separately. Obviously, the semi-classical approximation violates the symmetry between position and momentum. 

\begin{figure}[th]
    \centering   
    \includegraphics[width=6.5cm]{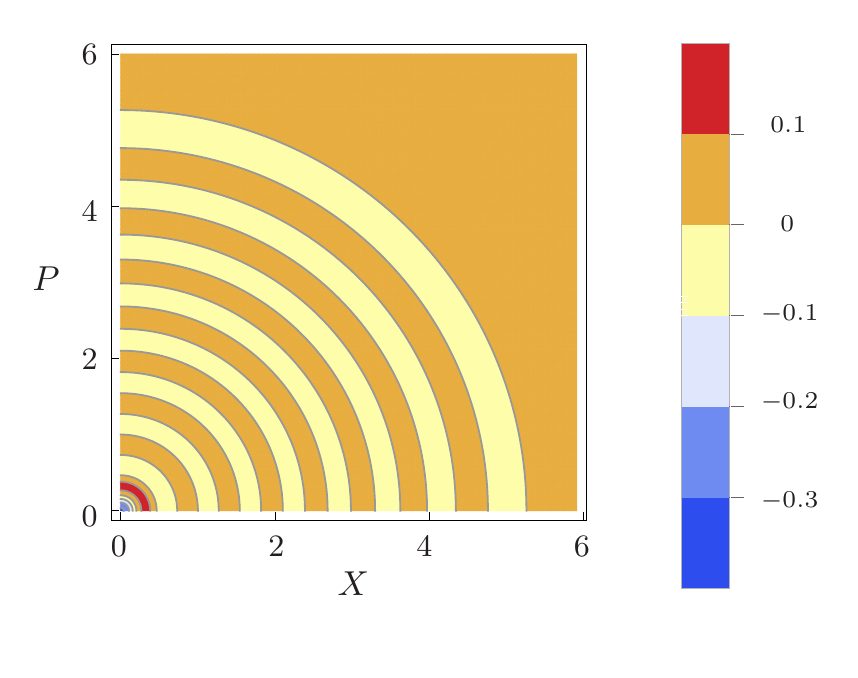}
     \includegraphics[width=6.5cm]{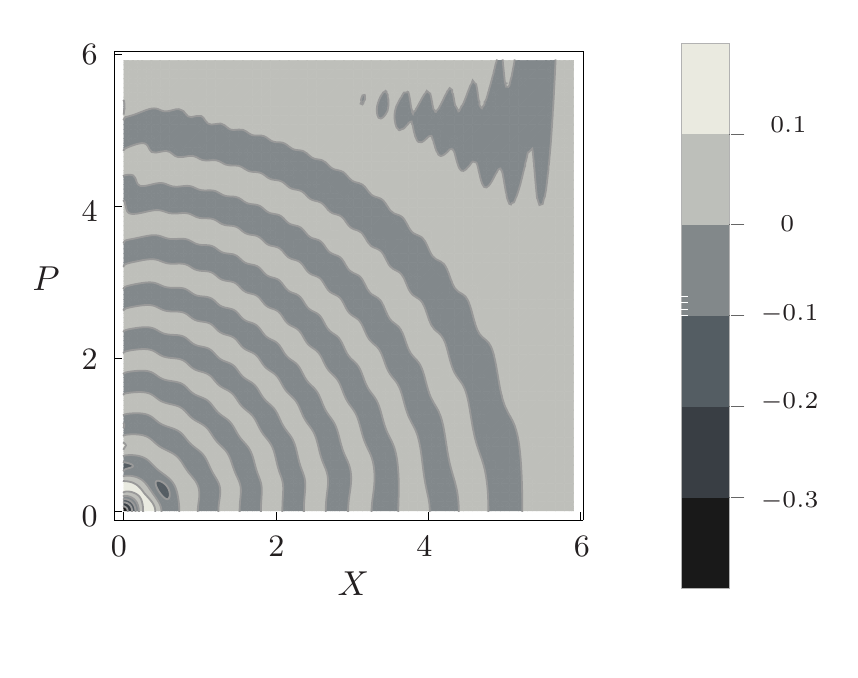}
    \caption{The exact Wigner function (top) and its semi-classical approximation analogue (bottom) obtained for \mbox{$n=17$}.}
    \label{fig:what}
\end{figure}

The Wigner function obtained from the semi-classical wave-functions is presented in Fig.~\ref{fig:what} and compared with the exact result.

\section{Classical limit}
Based on both Fig.~\ref{fig:fig1} and Fig.~\ref{fig:fig2} it is clear that the Wigner function exhibits oscillatory behavior even for large values of $n$ and does not resemble the classical distribution $w(X,P)$. A discussion of the classical limit requires, therefore, precise formulation of the way it is taken. {\it The~proper way} is to take the limit of $n \rightarrow\infty$ and $\hbar\rightarrow 0$ but with {\it $n \hbar \omega$ being a constant.} In other words, the total energy of the oscillator is equal to $E_{clas}$ and the energy difference between 'neighboring' states goes to zero. 

The classical limit should be understood in the following way. The Wigner function is a sort of distribution and it should tend to the limit in the sense of distributions. Namely, for every smooth function $f(x,p)$ and/or $f(X,P)$ the integral 
\begin{equation} 
\int \dd{}x\,\dd{}p\, W(x,p)f(x,p)\rightarrow \int \dd{}X\, \dd{}P\, w(X,P)f(X,P),
\end{equation} 
where $w(X,P)$ is the classical distribution of position and momentum, see Eq.~(\ref{classical distribution}). Any function $f(x,p)$ provides averaging over rapid oscillations of the Wigner function, so speaking about the classical limit is justified in this case. Physically, the function $f(x,p)$ can be understood as a result of finite resolution of the measuring device. Quantum effects are seen if the particle position is measured with accuracy  smaller than the typical de Broglie wavelength. In the limit of large $n$ the wavelengths become shorter than the spatial resolution of the measuring device and the classical limit emerges. 

There is also alternative interpretation. If the accuracy of measured energy is such that individual energy states cannot be distinguished the averaging over a range of energies is automatically performed. Contribution of the neighboring states to the Wigner function in such case leads to the classical distribution of position and momenta characteristic for the micro-canonical ensemble.

Behavior of the Wigner function can be understood from its semi-classical version given in~Eq.~(\ref{expansion2}). The Wigner function obtained from the approximate wave functions (in the WKB approximation) does not differ substantially from the exact function, except around the vicinity of turning points. Let us look at Eq.~(\ref{expansion2}). The term proportional to $\cos\left(2S(X)/\hbar\right)$ in the Wigner function, see~Eq.~(\ref{expansion2}) leads to oscillations in $X$ and $P$ variables. These oscillations, clearly seen in~Fig.~\ref{fig:fig1}, are present even for very large $n$. The~spatial period of these oscillations scales as $\frac{a}{\sqrt{n}}$. The scaling can be justified noticing that the wave-function, proportional to $\sin(S(X)/\hbar)$, has $n$ nodes extending to $X_{max}=\sqrt{\frac{2E}{m\omega^2}}=2^{3/2}a\sqrt{n}$. The average spatial wavelength (de Broglie wavelength) scales as $2X_{max}/n \simeq \frac{a}{\sqrt{n}}$.

Averaging over rapid spatial oscillations leads to elimination of the term proportional to $\cos(2S(X)/\hbar)$. Then, the Wigner function becomes:
\begin{eqnarray}
    \label{semiclassical wigner 0}
    &&W^{\rm wkb}(X,P)=  \\
    &&\frac{m}{T\, p(X)}\int\limits \frac{\dd{}s}{2\pi\hbar}\ \ee^{\ii\, sP/\hbar} \ \left( e^{-\ii\, s\, p(X)/\hbar} +
    \ee^{\ii\, s\, p(X)/\hbar}\right). \nonumber
\end{eqnarray}
We take the classical limit as before as the limit of small~$\hbar$, but keep the relevant physical quantity that has a classical meaning as a constant. This physical quantity is the energy $E_{clas}=n\hbar\omega $. Thus the limit is \mbox{$\hbar \rightarrow 0$,} $n\rightarrow \infty$ and $E_{clas}=n\, \hbar\omega = const$. 

Inspection of Eq.~(\ref{semiclassical wigner 0}) indicates that not all values of~$s$ contribute significantly to the integral. The important values are such that the local momentum $p(X)$ is approximately equal to the value of $P$. One can therefore extend the integration to the whole real axis, since only limited values of $s$ contribute. Hence,
\begin{equation}
    W^{\rm wkb}(X,P)= \frac{m}{T\, p(X)}\, \Big[\delta\big(P-p(X)\big)+
    \delta\big(P+p(X)\big)\Big]
\end{equation}
or even rewriting it accordingly 
\begin{equation}
    W_n^{\rm wkb}(X,P)=\frac{1}{T}\ \delta\left(\frac{1}{2m}\, P^2 +\frac{m\omega^2}{2}\, X^2  - E\right)\, .
    \label{limit}
\end{equation}
Observe that the Planck constant $\hbar$ is not present here, see also~\cite{Ripamonti}. Thus, Eq.~(\ref{limit}) has a classical meaning --- giving the classical distribution of $X,P$ position and momenta for the microcanonical ensemble parametrized by energy $E$. 

We have obtained the classical formula for the joined probability distribution for position and momentum, see the previous section.

\section{Conclusions}

The result given in Eq.(\ref{limit}) is the essence of these studies. It shows that all physical quantities expressed in terms of position and momentum have a~well-defined classical limit. The~classical~limit~is reached for small~$\hbar$, but at the same time the values of relevant physical quantities, like energy, are constant. In relation to states with definite energy, the classical limit of the Wigner function does not exist. An average over rapid spatial oscillations or average of Wigner functions over a reasonably broad energy range turns out to be the classical micro-canonical distribution. In this way, the probabilistic nature of quantum physics remains in the classical limit. 

\section*{Appendix I}

The relation between classical and quantum mechanics is sometimes presented with the help of action--angles variables. These variables are implicit in the semiclassical wavefunctions, the semiclassical energy quantization comes from the relation between the classical action $\int \dd{}y\, p(y)$, where the integral is taken over the whole period, see Eq.(\ref{classical action}), and the Planck constant
\begin{equation}
\int \dd{}y\, p(y)=2\pi\hbar n
\end{equation}
with $n$ being a natural number. 
We will now introduce the action - angle variables and show that the formula for $W(X,P)$ becomes even more elegant in action-angle variables, \cite{Landau}.  The action variable $I$ is defined as 
\begin{equation}
    I= \ointop \dd{}x\, p(x)
\end{equation}
where the integral is taken over the 
period. In case of harmonic oscillator the action variable is related to energy by a simple formula:
\begin{equation}
    E=I\omega
\end{equation}
and the angle variable $\varphi$, being a canonical coordinate canonically conjugate to $I$, is
\begin{equation}
    \varphi = \omega (t+t_0)
\end{equation}
The standard canonical variables $x$ and $p$ can be expressed in terms of $I$ and $\varphi$:
\begin{eqnarray}
x(I,\varphi)&=&\sqrt{\frac{2I}{m\omega}}\, \cos\varphi\nonumber\\
p(I,\varphi)&=&-m\, \omega \, \sqrt{\frac{2I}{m\omega}}\sin\varphi
\end{eqnarray}
The joined probability distribution of position and momentum can be expressed by integral over the angle variable:
\begin{eqnarray}
    w(X,P)&=&\frac{1}{2\pi}\int \dd{}\varphi\ \delta\big(X-x(t)\big)\ \delta\big(P-p(t)\big)\nonumber \\
    &&=\frac{1}{2\pi}\ \delta\big(I(X,P)-I\big).
    \label{classical distribution}
\end{eqnarray}

Thus the joined probability of position and momentum with the constant energy and random phase is concentrated around the chosen action with no phase dependence.

The case of harmonic oscillator is exceptional, the relation between energy $E$ and action variable $I$ is very simple, namely $E=\omega I$. Thus the change of variables from $E$ and $t_0$ to $I$ and $\varphi$ is almost trivial. This is, however, not the general case, in other system the transformation is much more complicated.

\section*{Appendix II}

We will now derive Eq.(\ref{Laguerre}) for the Wigner function related to an energy state $n$ of a harmonic oscillator. Let~us introduce the generating function of a complex number~$z$ for the energy states, namely
\begin{equation}
G(x,z)=\sum_{n=0}^{\infty}e^{-|z|^2/2}\, \frac{z^n}{\sqrt{n!}}\, \psi_n(x).
\end{equation}
For a coherent state $\psi_z(x)$, the generating function has the following form
\begin{equation}
    G(z,x)=\exp\left(-\frac{|z|^2+z^2}{2}+ \sqrt{2}\, z\,\frac{ x}{a}-\frac{x^2}{2a^2}\right).
\end{equation}
To obtain the wavefunction $\psi_n(x)$, one should expand $G(z,x)$ into a~power series in $z$, and selecting the terms proportional to \quad $e^{-|z|^2/2}\, \frac{z^n}{\sqrt{n!}}$.

In a similar manner the Wigner function $W_n(X,P)$ can be found. Let us now define the following auxiliary function 
\begin{eqnarray}
\label{generating Wigner}
  &K(&z_1,z_2,X,P) =  \\
  && \int \dd{}s \ G^*\left(z_1,X-\frac{s}{2}\right)\ \frac{\exp\left(\ii\, sP/\hbar\right)}{2\pi\hbar} \ G\left(z_2,X+\frac{s}{2}\right). \nonumber
\end{eqnarray} 
and apply the mentioned procedure. The integration in~(\ref{generating Wigner}) becomes easier, since this is a~gaussian integral. As a result, we get 
\begin{eqnarray}
&K&(z_1,z_2,X,P)=  
2\,  \exp{\left(-\frac{1}{a^2}X^2-\frac{a^2}{\hbar^2}P^2\right)} \nonumber\\
& \times & \exp\left(-\frac{|z_1|^2+|z_2|^2}{2}-z_1z_2^*\right)\qquad \qquad \nonumber \\
 & \times &\exp\left(2z_1\, \frac{X/a-\ii { Pa/\hbar}}{\sqrt{2}}+2z_2^*\, \frac{X/a+\ii{ Pa/\hbar}}{\sqrt{2}}\right).\nonumber \\
\end{eqnarray}
Next, expanding $K(z_1,z_2,X,P)$ into power series in $z_1$ and~$z_2$, and selecting the terms proportional to $(z_1^*z_2)^n$, we get
\begin{eqnarray}
    &W_n&(X,P)=\frac{(-1)^n}{\pi\, \hbar}\exp\left(-\frac{1}{a^2}X^2-\frac{a^2}{\hbar^2}P^2 \right) \nonumber \\ 
 & \times &\sum_{j=0}^n \left(\begin{array}{c}  n\\  j\\ \end{array}\right) \, \frac{(-1)^{j}\, 2^j}{j!}\, \left(\frac{1}{a^2}X^2+\frac{a^2}{\hbar^2}P^2\right)^{j}.
\end{eqnarray}
The sum is nothing else but the Laguerre polynomial $L_n\left(2\xi\right)$, where $\xi=\frac{1}{a^2}X^2+\frac{a^2}{\hbar^2}P^2$. This ends the derivation.
 


\begin{thebibliography}{99}

\bibitem{zurek} W.H. Zurek, ``Quantum Darwinism'', Nature Physics 5, 181 (2009)

\bibitem{Landau}
L.D. Landau, E.M. Lifshitz, \textit{Quantum mechanics - nonrelativistic theory} (Pergamon Press, Cambridge, 2017).

\bibitem{Klauder}
J.R. Klauder and B. Skagerstam, ``Coherent States'', World Scientific, Singapore, 1985,

\bibitem{Glauber}
Roy J. Glauber, Coherent and Incoherent States of the Radiation Field, Physical Review 131, 2766 (1963),
doi:10.1103/physrev.131.2766

\bibitem{Schleich} W. Schleich, Quantum Optics in Phase Space (Wiley-VCH, Berlin, 2001).

\bibitem{Case} William B. Case, Wigner functions and Weyl transforms for pedestrians,  Am. J. Phys. 76, 937,  (2008)

\bibitem{textbook}See for example: C. Zachos, D. Fairlie, T. Curtright: Quantum mechanics in phase space: an overview with selected papers, World Scientific, 2005 ISBN 978-981-4520-43-0

\bibitem{Hillery84} M. Hillery, R. F. O’Connell, M. O. Scully, and E. P. Wigner, ``Distribution functions in physics: Fundamentals'', Phys. Rep. 106, 121–167 (1984).

\bibitem{Lee} H. Lee, “Theory and application of the quantum phase-space distribution functions,” Phys. Rep. 259, 150–211 (1995).

\bibitem{Tatarskij83}
V. I. Tatarskij, ``The Wigner representation of quantum mechanics'', Sov. Phys. Uspekhi 26, 311, (1983), https://doi.org/10.1070/PU1983v026n04ABEH004345

\bibitem{Ripamonti}
N Ripamonti, Classical limit of the harmonic oscillator Wigner functions in the Bargmann representation 1996, Journal of Physics A: Mathematical and General, Volume 29, Number 16, 1996

\bibitem{Wang}
Jianhua Wang, Kang Lib, and Sayipjamal Dulat, Wigner Functions for harmonic oscillator in noncommutative phase space,  arXiv:0908.1703v1

\bibitem{Wigner2}
E. P. Wigner, ``On the quantum correction for thermodynamic equilibrium'', Phys. Rev. 40, 749-759 (1932)

\end{thebibliography}
\end{document}